# Spin-Valley Relaxation of Rydberg Excitons

V. Jindal[1*], K. Mourzidis[1*], M. Semina[2], D. Lagarde[1], A. Balocchi[1], P. Renucci[1], T. Boulier[1], T. Taniguchi[3], K. Watanabe[4], M. Glazov[2†] and X. Marie[1,5†]

[1] *Université de Toulouse, INSA-CNRS-UPS, LPCNO, 135 Avenue Rangueil, 31077 Toulouse, France*
[2] *Ioffe Institute, 26 Polytechnicheskaya, 194021 Saint Petersburg, Russia*
[3] *International Center for Materials Nanoarchitectonics, National Institute for Materials Science, 1-1 Namiki, Tsukuba 305-00044, Japan*
[4] *Research Center for Functional Materials, National Institute for Materials Science, 1-1 Namiki, Tsukuba 305-00044, Japan*
[5] *Institut Universitaire de France, 75231 Paris, France*

*Rydberg excitons, characterized by large spatial extension and reduced electron-hole overlap, must have a spin-valley dynamics different from that of ground state excitons. Here we report a direct measurement of spin relaxation of Rydberg excitons in high-quality WSe$_2$ monolayer using continuous-wave and time-resolved optical orientation experiments. Excited excitonic states exhibit exceptionally large photoluminescence circular polarization, approaching 90% for the 3s state. Time-resolved measurements reveal a strong increase of the spin relaxation time with the principal quantum number, from ~2 ps for the 1s exciton to ~75 ps for the 3s exciton. A microscopic model based on electron-hole exchange-driven spin relaxation quantitatively reproduces the observed trend, demonstrating that Rydberg excitons enable tunable spin–valley dynamics in two-dimensional semiconductors.*

* These authors contribute equally to this work
† corresponding authors : [marie@insa-toulouse.fr](marie@insa-toulouse.fr), [glazov@coherent.ioffe.ru](glazov@coherent.ioffe.ru)

Excitons - Coulomb-bound electron–hole pairs - dominate the optical response of semiconductors and play a central role in spin and valley physics in reduced dimensions [1–3]. Their internal structure reflects the interplay of band dispersion, dielectric screening, and spin–orbit coupling, and their dynamics underpin numerous optoelectronic and spin/valley functionalities. In particular, the exciton spin degree of freedom, accessed optically through circular polarization, is a key resource for encoding and manipulating information on quantum level; its relaxation time sets the fundamental limit for optical orientation and coherent control.

Rydberg excitons are highly excited electron–hole pairs whose internal structure mirrors that of atomic Rydberg states. Their large spatial extent, strong interactions, and large optical nonlinearities make them uniquely sensitive probes of many-body and quantum-optical phenomena in solids [4]. Rydberg excitons are also very sensitive detectors of strongly correlated electronic phases [5–7]. They have been evidenced in various semiconductor structures. Among these, the most extensively studied system is cuprous oxide ($Cu_2O$), where excitons with principal quantum numbers as high as $n = 30$ have been observed [8,9]. In most other bulk or low-dimensional semiconductors, however, excited exciton states beyond $n = 4$ are rarely resolved because of line broadening, disorder, and rapid relaxation [10,11]. These constraints have strongly limited the exploration of dynamical properties—including spin relaxation—in the genuine Rydberg regime, despite the growing interest in leveraging Rydberg excitons as solid-state analogues of atomic Rydberg systems.

The spin relaxation of ground state exciton has been intensively studied in bulk semiconductors, quantum wells, and more recently in two-dimensional materials. In monolayer (ML) transition-metal dichalcogenides (TMDs), the interplay between the inversion symmetry breaking and the spin-orbit interaction leads to a unique coupling of the charge carrier spin and the *k*-space valley physics [12]. The circular polarization of the absorbed or emitted photon can be directly associated with selective carrier excitation in one of the two non-equivalent K valleys [13–15]. The strong Coulomb interaction and the spin–valley locking enable efficient optical initialization of exciton spin-valley polarization, and a large body of theory and experiment has established the microscopic relaxation channels. However, this progress has focused almost exclusively on the exciton ground state, usually the bright 1s exciton [16–19]. The spin properties of excited, Rydberg-like excitons in 2D systems remain largely unexplored, despite the advances in observing narrow high-lying exciton series in high-quality TMD MLs [10,20–23]. These Rydberg excitons are expected to offer a qualitatively different regime for spin dynamics. It has been shown that the exciton spin relaxation in TMDs is mainly controlled by the electron–hole exchange interaction [18,19,24–26]. This exchange energy being proportional to the electron-hole overlap decreases with the principal quantum number $n$ since the exciton radius expands. Thus, one can anticipate a substantial slowing down of spin-valley relaxation for excited exciton states. At the same time, the scattering time that governs exchange-driven depolarization is also expected to evolve with $n$. Quantifying how these competing trends shape the spin lifetime across the Rydberg series is essential both for a microscopic understanding of exciton spin physics and for assessing the potential of highly excited excitons as long-lived spin-valley species.

In TMD MLs, larger exciton luminescence polarization was reported in continuous wave (*cw*) optical orientation experiments for *n*=2 excited state (2s) compared to *n*=1 in a few studies [20,27]. However, larger polarization of the excited states luminescence does not necessarily imply longer spin-valley relaxation times $\tau_s$ (for simplicity, "spin-valley" is hereafter denoted as "spin"). Let us recall that in a simple approach, the measured polarization *P* in stationary conditions writes as [1] :
$$P=P_0/(1+\tau/\tau_s), \tag{1}$$
where $P_0$ is the maximum polarization of photoexcited species linked to the selection rules and exciton formation processes. As the lifetime τ of the excited Rydberg states is typically shorter due to relaxation to the ground state, the larger polarization of *n*=2 excited state could simply be a result of faster population decay from the excited state rather than a true increase of spin relaxation time $\tau_s$. To distinguish between these possibilities, time-resolved optical orientation experiments are essential. Such experiments allow direct measurements of the spin relaxation dynamics [1] and provide insights into whether the larger polarization is indeed a consequence of slower spin relaxation in the excited state, or merely due to the differences in the exciton lifetimes. However, performing such time-resolved measurements remains technically challenging, as the Rydberg excitons exhibit very short lifetimes.

In this Letter, we report on measurements and interpretation of the spin dynamics of Rydberg excitons in high-quality WSe$_2$ monolayers. Using continuous-wave and time-resolved photoluminescence (PL), we directly access the spin/valley polarization dynamics of excited excitons. In polarized *cw* photoluminescence we observe an exceptionally large polarization for the $n = 3$ (3s) exciton, reaching ∼ 90%, *i.e.* about three times larger than the one observed for the 1s exciton. Time-resolved optical orientation experiments reveal that this enhancement arises from a dramatic increase of the spin relaxation time $\tau_s$ with $n$: we measure $\tau_s \approx 75$ ps, 35 ps, and 2 ps for $n = 3, n = 2$, and $n = 1$ excitons, respectively. To account for these observations, we develop a model of exchange-driven spin relaxation that explicitly incorporates the $n$-dependence of both the exciton exchange interaction and the exciton momentum-scattering time. The latter is dominated by exciton–acoustic-phonon coupling in our samples. The theory reproduces the measured hierarchy of spin lifetimes across the Rydberg series in good quantitative agreement, establishing Rydberg excitons in monolayer TMDs as a new platform where spin relaxation can be tuned over more than an order of magnitude by the internal exciton quantum number.

The investigated samples are WSe$_2$ MLs encapsulated in hexagonal boron nitride (hBN) and deposited onto a SiO$_2$/Si substrate using a dry-stamping technique (Fig. 1(a)) [28,29]. Time-resolved and *cw* micro-photoluminescence experiments are performed at 5 K with a vibration-free closed-cycle cryostat. Details on the samples and experimental setups can be found in the Supplemental Material (SM) [30], which includes Refs [31,32]. Figure 1(b) presents the *cw* photoluminescence spectrum together with the differential reflectivity. The spectra are dominated by the intense signal corresponding to the ground state 1s exciton with a narrow linewidth of ∼ 2 meV (Full Width at Half Maximum, FWHM). Remarkably, we also observe clearly the signatures of exciton excited states *n*s up to *n*=4 in hot photoluminescence and *n*=5 in the reflectivity spectrum [10,33]. The binding energy of the 1s exciton is about 170 meV, as determined by magneto-absorption experiments performed in a similar sample [22,34]. The exciton Rydberg series do not follow the well-known hydrogenic series as a consequence of the inhomogeneous dielectric environment [3,35]. Figure S1 in the Supplemental Material shows that the binding energy of the different excited states are well reproduced by a model based on the Rytova-Keldysh potential [30,36].

To probe the exciton spin properties, we have performed *cw* optical orientation experiments: the black and red curve in Fig. 2 display the right (I$^+$) and left (I$^-$) circularly-polarized luminescence components following σ$^+$ polarized excitation laser. The steady-state circular polarization degree P$_c$= (I$^+$ - I$^-$)/(I$^+$ + I$^-$) is also plotted. For the exciton ground state 1s, we measure P$_c$ = 33 ± 1 %, in agreement with the previous reports [17,37]. The striking feature is that for the exciton excited states polarization increases drastically with *n*, reaching values up to 80± 1 % for *n*=2 and even 90 ± 2 % for *n*=3 (measurements performed on a second sample show very similar results, Fig. S2 in the SM [30]) . This demonstrates that the exciton spin relaxation time $\tau_s$ for these excited states is longer than their lifetime τ. Interestingly, the measured polarization of the exciton states does not depend on the excitation power (in the range 1-50 µW) as shown in Fig. 2(b). Importantly, we obtain the similar trends for the exciton spin-valley coherence: the PL linear polarization of the Rydberg excitons following linearly-polarized excitation also increases with *n* (Fig. S3 in the SM) [30].

To interpret the circular polarization rates measured in steady state for the different excitonic states, it is essential to perform time-resolved measurements to disentangle the respective roles of spin relaxation time and excited state lifetime, Eq. (1). Figure 3(a) presents the time evolution of I$^+$ and I$^-$ PL components of the 3s exciton following a σ$^+$ polarized picosecond excitation. The optimized time-resolution allows us to measure the 3s exciton lifetime $\tau^3$ ~1.6± 0.2 ps (note that the PL decay time is clearly longer than the detected laser pulse scattered from the sample surface, Fig. S4 in the SM) [30]. The initial polarization P(t=0) is very large, about 90%, demonstrating that the energy relaxation and the exciton formation process yield negligible depolarization [38]. We also plot in Fig. 3(a) the time evolution of the polarization degree. Using a simple mono-exponential fit, we get a spin decay time $\tau_s^3 = 75 \pm 35$ ps, one order of magnitude longer than the spin-valley relaxation time reported for 1s exciton [16,39]. Figure 3(b) displays the circular polarization dynamics for the different exciton states $n$s. The increase of the exciton spin relaxation time as a function of $n$ is clearly observed as we measure $\tau_s^1 = 2.2 \pm 0.3$ ps, $\tau_s^2 = 35 \pm 10$ ps and $\tau_s^3 = 75 \pm 35$ ps for $n$=1, 2 and 3 states respectively. This trend is in good agreement with the spin polarization measured in stationary conditions (Fig. 2).

In addition to the depolarization time, the time-resolved photoluminescence experiments also yield a direct measurement of the lifetime of the exciton excited states $\tau^n$. Figure S4 in the Supplemental Material displays the corresponding kinetics and we find $\tau^1$=1.9±0.2 ps, $\tau^2$=1.5 ± 0.2 ps and $\tau^3$=1.6 ±0.2 ps for the $n$=1, 2 and 3 states respectively [30]. These results demonstrate that the lifetime of the Rydberg exciton states is not controlled by their radiative recombination. Indeed the radiative recombination time $\tau^n$ of hydrogen-like 2D $n$s excitons should follow a scaling law $\tau^n = (2n-1)^3 \cdot \tau^1$ [40,41]. This means that the radiative recombination time of the $n$=2 exciton for instance should be at least one order of magnitude longer than the one of 1s exciton [29,42], even if we consider the deviation from the hydrogenic 2D model considering the Rytova-Keldysh potential. The measurement of a decay time of ~ 1.5 ps for $n$=2 and 3 states demonstrates that their lifetime is not radiative; remarkably our measured decay times are in excellent agreement with calculations based on a fully quantum mechanical description of momentum- and energy-resolved exciton dynamics [43]. In contrast to Cu$_2$O, the Rydberg excitons lifetime in WSe$_2$ MLs is controlled here by the relaxation to the ground state $n = 1$ [4,44]. As the measured exciton lifetime does not significantly vary with $n$, the spectacular increase of the *cw* polarization as a function of $n$ in Fig. 2 is indeed a direct consequence of the increase of the spin relaxation time as probed directly in the time-resolved measurements presented in Fig. 3. Estimates based on Eq. (1) and experimental values of $\tau_s^n$ and $\tau^n$ confirm this conclusion.

We have developed a model to calculate the spin relaxation of the excitons with principal quantum number $n$ larger than 1. Let us recall that the main spin relaxation mechanism for 1$s$ excitons in quantum wells and TMD monolayers is related to the electron-hole exchange interaction, which results in a simultaneous flip of both electron and hole spins [18,19,25,45]. In the collision dominated regime, the corresponding spin relaxation time of $n$th Rydberg state $\tau_s^n$ writes simply [18]:

$$\frac{1}{\tau_s^n} = \langle (\Omega_K^n)^2 \cdot \tau_n^* \rangle, \qquad (2)$$

where $\Omega_K^n = \alpha_n K$ is the precession frequency corresponding to the long-range exchange interaction of the excitons with the center-of-mass wavevector $K$, $\alpha_n$ is the exchange-interaction parameter related to the exciton radiative decay rate [18], $\tau_n^*$ is the exciton scattering time and the angular brackets denote averaging over the exciton energy distribution [18].

In WSe$_2$ monolayers characterized by a very large exciton binding energy, and as a consequence very significant characteristic exchange energy, Eq. (2) is inapplicable for the 1s exciton because the collision-dominated regime is not realized in this case, since $\Omega_K^1 \tau_1^* \gg 1$ for typical exciton wavevectors. The detailed analysis of the 1s exciton time-resolved polarization (which will be reported elsewhere) yields a spin relaxation time of the the 1s exciton of the order of ~ 1 ps in agreement with the literature [39,46,47].

To evaluate the spin relaxation of the exciton excited states ($n$>1), we have calculated the dependence of both the exchange energy and the scattering time (the two key parameters in Eq. (2)) as a function of exciton principal quantum number $n$. As shown below, the conditions of the collision-dominated regime ($\Omega_K^n \tau_n^* \ll 1$) are approximately verified for $n > 1$ states and Eq. (2) can be used for qualitative analysis. The binding energies of the Rydberg states are progressively smaller resulting in larger spatial extent of the exciton wavefunctions. It results in a strong decrease of the exchange interaction $\alpha_n$ potentially yielding the increase of $\tau_s^n$ in Eq. (2).

Specifically, the exchange interaction variation can be estimated from the change of the radiative broadening $\Gamma_n$ as $\alpha_n \propto \Gamma_n$ [18]. We have calculated the ratio of the radiative broadening $\Gamma_n$ of the exciton n with respect to the one of the exciton ground state 1s. The result is plotted in Fig. 4(a): We predict $\frac{\alpha_2}{\alpha_1} \approx 7$ and $\frac{\alpha_3}{\alpha_1} \approx 17$, as a result of the increase of the radiative lifetime due to the decrease of the electron-hole overlap with *n*. These calculated ratios are very similar to the ones which can be evaluated from the fit of the measured reflectivity spectrum (see Fig. 4(a) and the detailed fits in Fig. S5 in the SM [30]). According to Eq. (2), this effect of the exchange interaction alone should result in ~ 50-fold and ~300-fold increase in the spin relaxation time for *n*=2 and *n*=3 excitons respectively, which is not observed in experiment.

However, the determination of the exciton spin relaxation time also requires calculations of (i) the exciton distribution over the wavevectors and (ii) the exciton scattering time $\tau_n^*$ for the different Rydberg states, Eq. (2). We stress that the measured time-resolved circular polarization degree corresponds to the spin dynamics of a small fraction excitons that remain in the Rydberg states on the timescale that exceeds their average lifetime $\tau^n$. In the time domain considered here (the first ~10 ps following photogeneration), we assume that the scattering time is dominated by the interaction with long-wavelength acoustic phonons (see the schematics in Fig. 4(b) illustrating the dynamics for n=3 excitons). Indeed, Monte Carlo calculations of the 1s exciton energy relaxation in TMD MLs following non-resonant excitation have shown that the initial drop of kinetic energy is dominated by the emission of optical phonons and ends typically ~1 ps after photoexcitation [48]. After this ultra-fast cooling which cannot be resolved in our ps experiments, the excitons interact mainly with acoustic phonons and slowly thermalize with the lattice in at least 20-30 ps, see schematics in Fig. 4(b) [49] and SM [30] for details. As the exciton spin relaxation of the different Rydberg states occur in the first tens of ps, we consider that the main scattering mechanism involves these acoustic phonons.

The interaction between excitons and acoustic phonons in TMDs is dominated by the deformation potential contribution [50]. For an exciton with a principal quantum number *n*, the scattering from a state with wavevector *K* to a wavevector *K'* accompanied by the emission or absorption of an acoustic phonon with wavevector q can be described by the following matrix element:

$$M_{KK'}^n = \sqrt{\frac{\hbar}{2\rho\Omega_q S}} q(D_c - D_v)\mathcal{F}^n(q)\delta_{K,K'+q}, \qquad (3)$$

where ρ is the mass 2D density of the TMD ML, $\Omega_q = sq$ is the phonon frequency, *s* is the sound velocity, *S* is the normalization area, $D_c$ and $D_v$ are the deformation potential constants for the conduction and valence band respectively. The form factor for the exciton with the principal quantum number *n* writes:

$$\mathcal{F}^n(q) = \int e^{-\frac{iq\rho}{2}} |\varphi^n(\rho)|^2 d\rho, \qquad (4)$$

where $\varphi^n(\rho)$ is the wave function of s-like 2D exciton state and we assumed for simplicity equal electron and hole effective masses.

The form-factor at zero wave vector $\mathcal{F}^n(0) \equiv 1$ for any *n*, it decreases with increase in *q* for small wavevectors; the decrease is steeper, the larger *n* is because the exciton spatial extension in Rydberg states increases with *n*, see SM for details [30]. To illustrate this dependence, we plot in Fig. 4(c) $(\mathcal{F}^n)^2$ for *n* varying between *n*=1 and 6 (calculated for the thermal wavevector $K_T = \sqrt{2mk_BT/\hbar^2}$ at the exciton temperature *T*=5, 15 and 30 K, *m* is the exciton translational mass). As a result, the exciton-phonon scattering for the excited states is reduced as compared to the ground state 1s [51]. According to Eq. (2), this leads to a decrease of the spin relaxation time, *i.e.* a dependence on *n* that is opposite to the one already discussed for the exchange energy. As shown below it partially compensates the decrease in $\Omega_K^n$ related to the decrease of the electron-hole overlap.

Crucially, on the $\tau^n \sim 1$ ps timescale, only a small fraction of Rydberg excitons reaches the light cone and radiatively recombines, thereby contributing to the observed circular polarization. In this time window, the energy exchange of excitons with the lattice in the energy range of $\sim k_B T$ is dominated by the long-wavelength acoustic phonons ($T = 5 \ldots 30$ K is the exciton temperature and $k_B$ is the Boltzmann constant); it occurs on the time scales $\tau_\varepsilon^n \gg \tau_n^*, \tau_{sv}^n$, see SM [30] for details. Hence, the observed PL polarization is dominated by the fraction of excitons with small energies which can reach the light cone in one exciton-phonon scattering act. Thus, we evaluate the pseudospin dynamics $S_z^n(t; E_K) \propto \exp[-t/\tau_s^n(E_K)]$ at a given exciton kinetic energy $E_K$ and calculate the spin dynamics as

$$S_n(t) = \frac{\int dE_K \, S_z^n(t; E_K) f(E_K)/\tau_\varepsilon^n(E_K)}{\int dE_K \, f(E_K)/\tau_\varepsilon^n(E_K)}, \qquad (5)$$

where $f(E_K)$ is the Boltzmann distribution function describing exciton occupancies and the factor $1/\tau_\varepsilon^n(E_K)$ is proportional to the probability of exciton to reach the light cone. The dynamics in Eq. (5)

were fitted with exponential function to determine the spin relaxation times, as described in SM [30]. Corresponding times for $n = 1, \ldots, 6$ calculated at $T = 10$ K are shown in Fig. 4(d) together with experimental values. Overall, we do observe the increase of the spin relaxation time with n and the agreement with experimental results is good despite the simplified approach. A full modeling of Rydberg exciton kinetics that includes thermalization and energy relaxation processes could be developed in the future.

In summary, we have demonstrated that Rydberg excitons in monolayer WSe$_2$ exhibit a pronounced slowing down of spin–valley relaxation with increasing principal quantum number. Time-resolved optical orientation measurements reveal that the spin relaxation time increases by more than an order of magnitude from the ground-state exciton to *n*=3 excited states. We develop a microscopic model of exchange-driven spin relaxation that incorporates the *n* dependence of both the long-range electron–hole exchange interaction and exciton–phonon scattering. The theory quantitatively reproduces the observed hierarchy of spin lifetimes and identifies the competition between reduced exchange splitting and weakened acoustic-phonon scattering as the key mechanism. Our results identify Rydberg excitons as a previously unexplored regime of exciton spin physics and open new opportunities for engineering long-lived spin and valley degrees of freedom in two-dimensional semiconductors.

**Acknowledgments**

This work was supported by the Agence Nationale de la Recherche under the program ESR/EquipEx+ (Grant No. ANR-21-ESRE- 0025), the France 2030 government investment plan managed by the French National Research Agency under Grant Reference No. PEPR SPIN ANR-22-EXSP0007 (SPINMAT) and the ANR project SOT-SpinLED. We also acknowledge support by the EU-funded DYNASTY project, ID: 101079179, under the Horizon Europe framework program. We thank C. Robert for his advice in the fabrication of one of the samples.

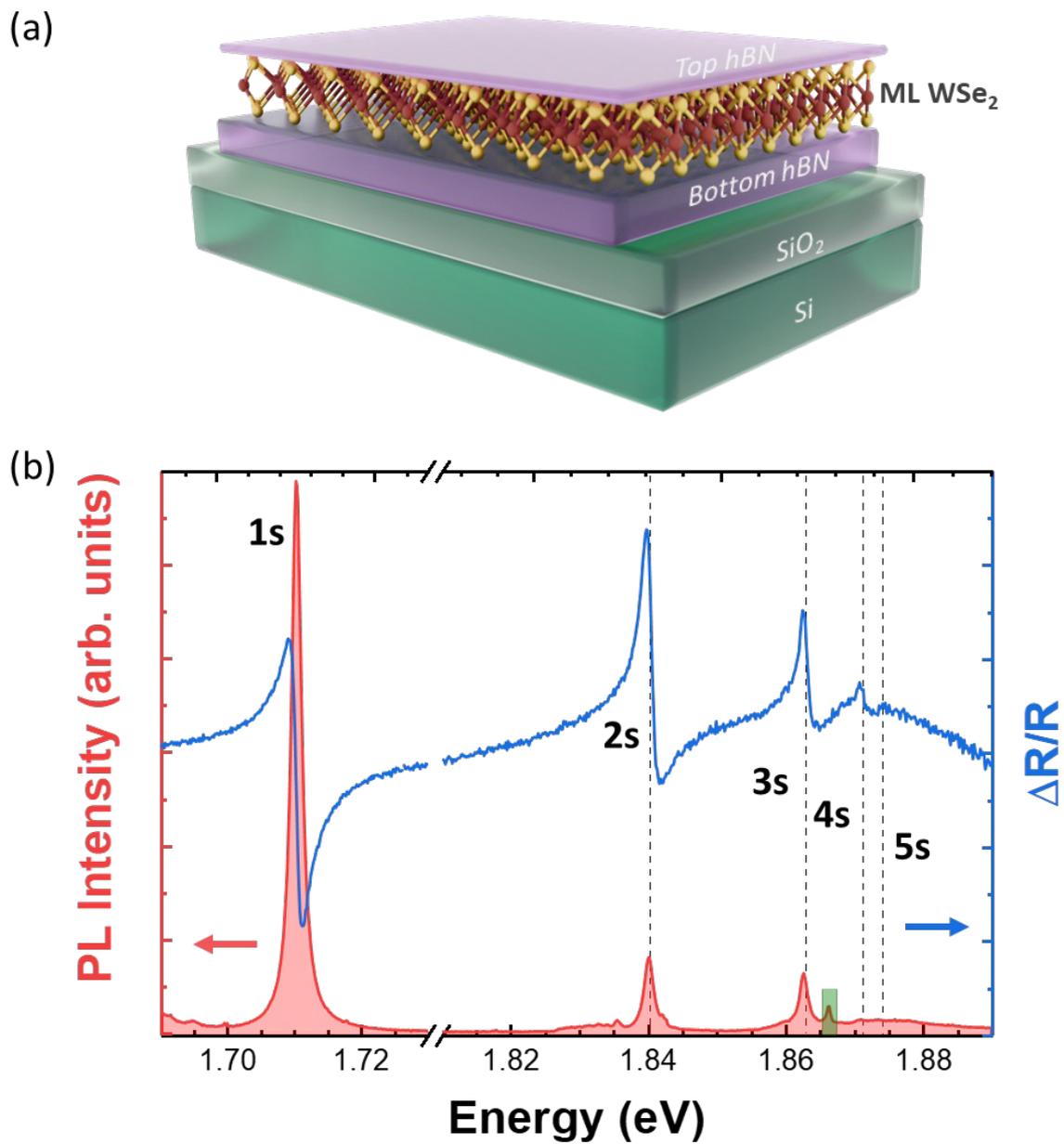

Figure 1. **(a)** Schematics of the WSe$_2$ monolayer; **(b)** Differential reflectivity and cw photoluminescence spectra showing Rydberg exciton states up to n=5, for the PL measurement, the laser excitation energy is 1.96 eV (the green area corresponds to a Raman peak).

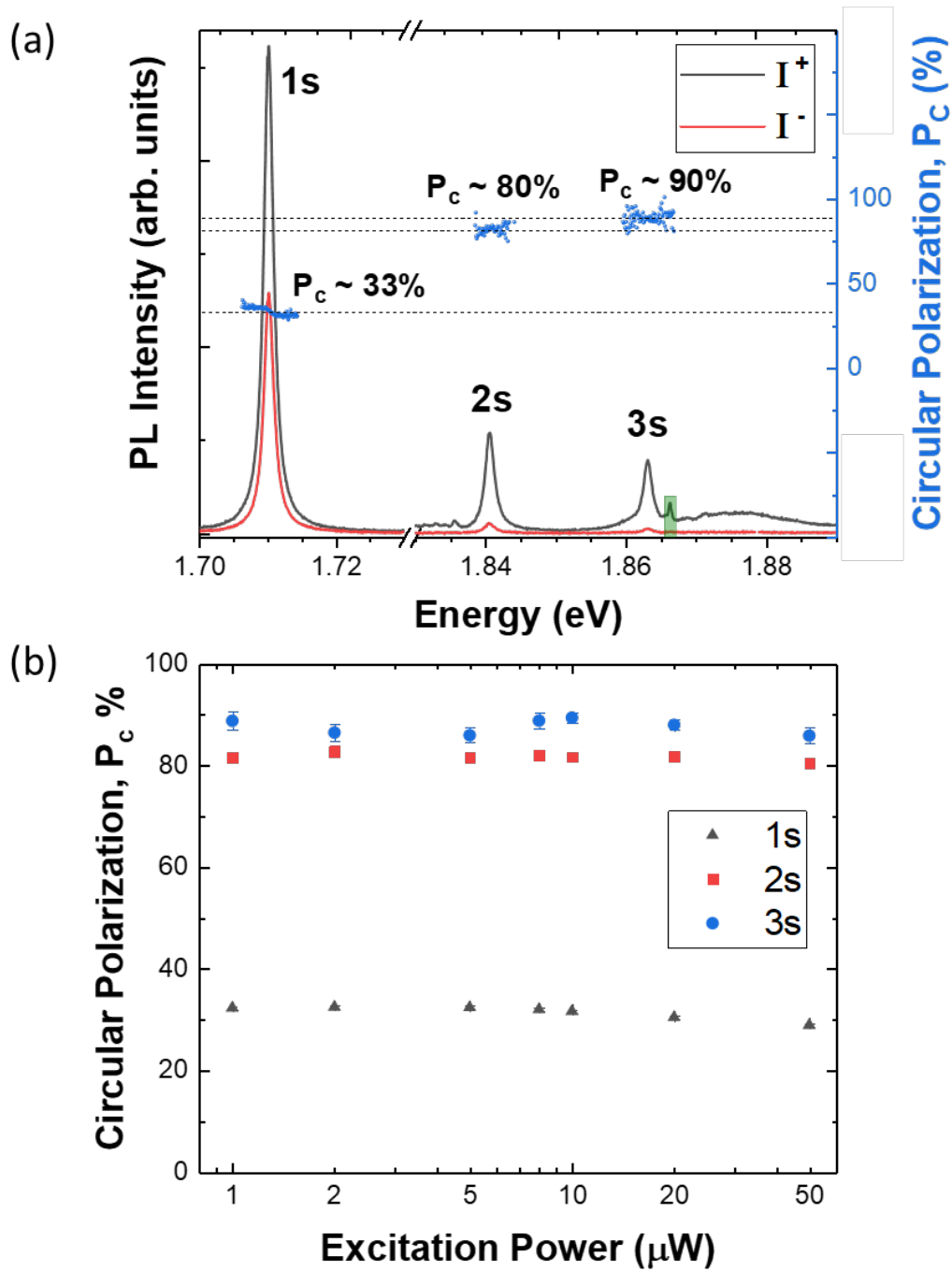

Figure 2 **(a)** Right ($I^+$) and left ($I^-$) circularly polarized luminescence following right circularly-polarized laser ($E_{exc}$=1.96 eV); the polarization degree $P_c$ of the different exciton states n=1 to 3 is also plotted; **(b)** PL circular polarization of the exciton states *n*s as a function of excitation power.

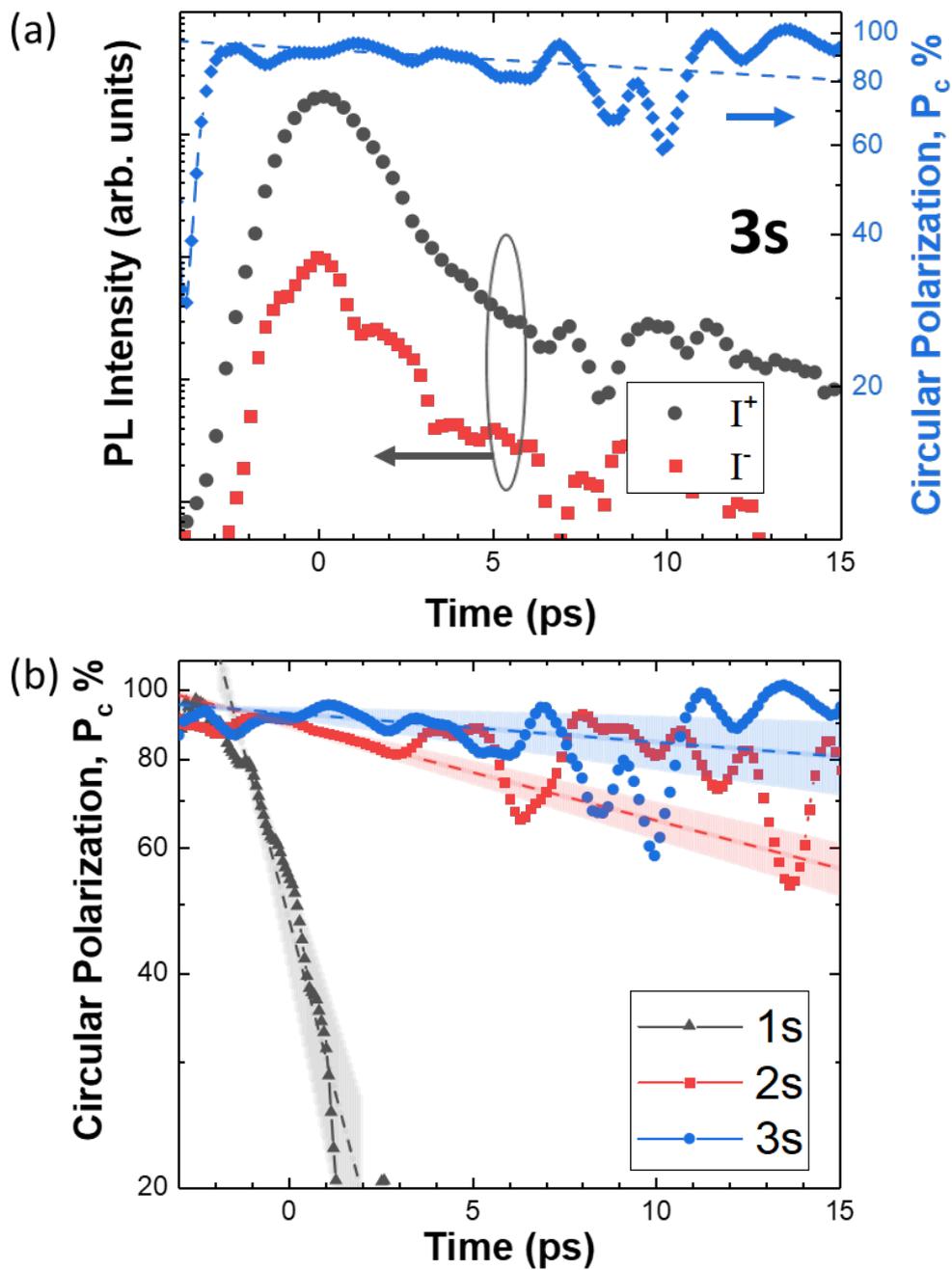

Figure 3 **(a)** Time evolution of the polarized luminescence components and the corresponding circular polarization $P_c$ following a right circularly-polarized σ+ picosecond laser pulse ($E_{exc}$=1.91 eV); **(b)** Circular polarization dynamics of the exciton states ns.

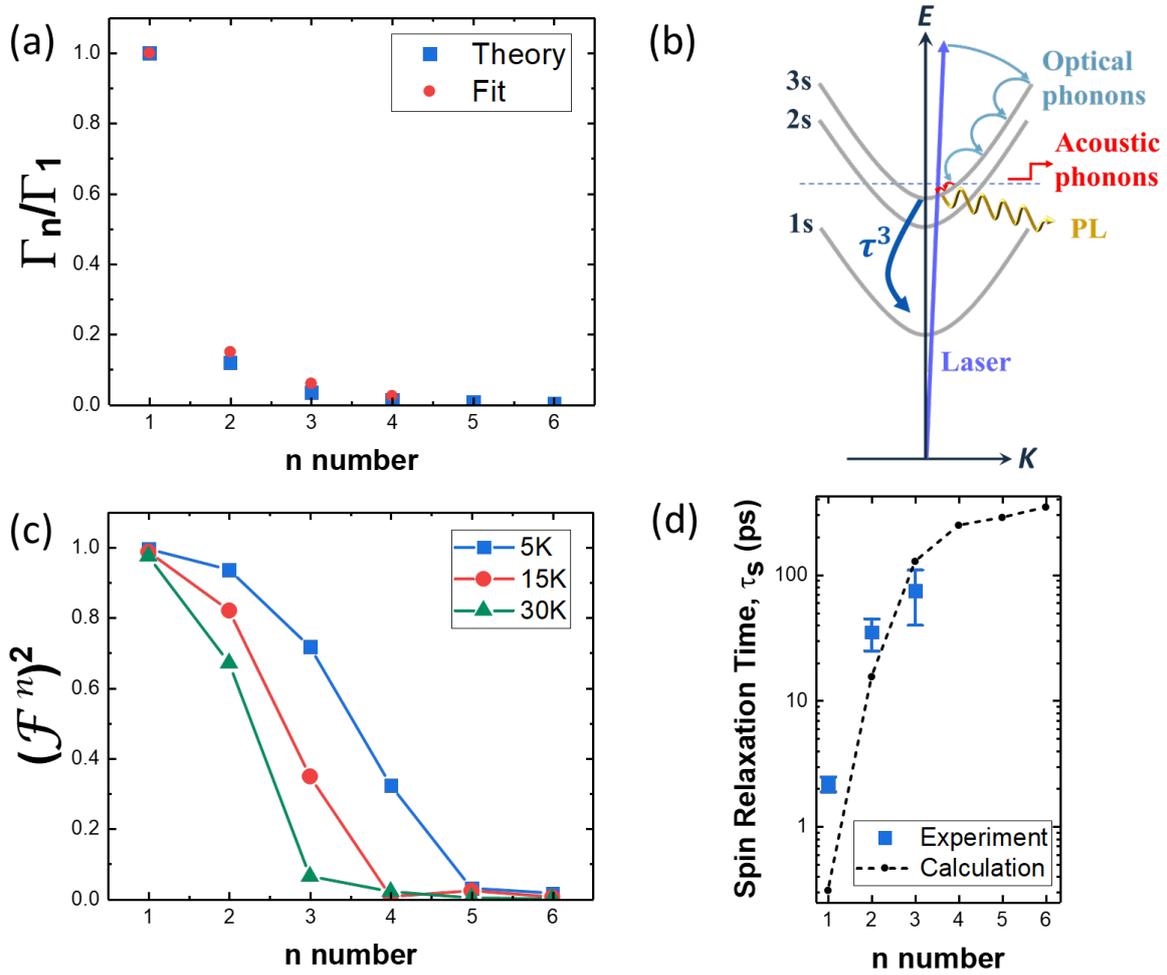

Figure 4 **(a)** Ratio of the radiative broadening $\Gamma_n$ of exciton n to that of the ground state 1s ; the blue squares correspond to the calculated value whereas the red circles are obtained from the fit of the measured reflectivity spectrum shown in Fig. 1(b), see text. **(b)** Schematic illustration of the exciton series dispersion curve, showing the main relaxation channels of exciton excited states involving optical and acoustic phonons (see text); for the sake of clarity, we just highlight here the dynamics of 3s excitons characterized by a lifetime $\tau^3$. **(c)** Calculated form factor $(\mathcal{F}^n)^2$ for n=1 … 6 (effective exciton temperatures T=5, 15 and 30 K); this form factor controls the dependence of the exciton *n*s-acoustic phonon interaction as a function of *n*. **(d)** Calculated (line connecting black dots) and measured ( blue symbols) exciton spin relaxation time as a function of n.